\documentclass[acus]{JAC2000}

\usepackage{graphicx}

\begin{document}
\title{\flushright{T10}\\[15pt] \centering 
SND UPGRADE}
\author{
G.N.Abramov,
M.N.Achasov,
V.M.Aulchenko,
K.I.Beloborodov, 
A.V.Berdyugin,\\
A.G.Bogdanchikov,
A.V.Bozhenok,
D.A.Bukin,
M.A.Bukin,
S.V.Burdin,\\
A.F.Danilyuk,
T.V.Dimova,
A.A.Drozdetsky,
V.P.Druzhinin\thanks{e-mail: druzhinin@inp.nsk.su},\\
V.B.Golubev,
V.N.Ivanchenko,
P.M.Ivanov,
A.A.Korol,
S.V.Koshuba,\\
E.A.Kravchenko,
L.V.Maksimov,
A.E.Obrazovsky,
A.P.Onuchin,
A.V.Ovchar,\\
E.V.Pakhtusova,
M.A.Peshkov,
V.M.Popov,
E.E.Pyata,
A.A.Salnikov,\\
S.I.Serednyakov,
V.V.Shary,
Yu.M.Shatunov,
Z.K.Silagadze,
A.A.Sirotkin,\\
A.G.Skripkin,
Yu.V.Usov,
A.V.Vasiljev,
Yu.S.Velikzhanin
\vspace{1mm}\\
Budker Institute of Nuclear Physics \\
630090, Novosibirsk, Russia}

\maketitle

\begin{abstract}
The program of upgrade of the Spherical Neutral Detector
for future experiments  at a new VEPP-2000 $e^+e^-$ collider
is presented. Modernization includes upgrades of
electromagnetic calorimeter, tracking system,
detector electronics, data acquisition system, and offline software.

It is also planned to equip the detector with two new subsystems:
particle identification system based on aerogel \v{C}erenkov counters and
external electron tagging system for $\gamma\gamma$ physics.
\end{abstract}

\section{Introduction}

The Spherical Neutral Detector \cite{snd} was designed for experiments
at VEPP-2M $e^+e^-$ collider in the energy range from 0.36
to 1.4 GeV. These experiments were carried out since 1996
till 2000. The SND collected about 30 pb$^{-1}$ of integrated
luminosity which corresponds to 7, 4 and 20 millions produced $\rho$,
$\omega$ and $\phi$ mesons, respectively. The main physical
goals of SND were precise measurements of the magnetic dipole
decays of $\rho$, $\omega$ and $\phi$, observation and study
of the electric dipole decays $\phi\to a_0\gamma, f_0\gamma$,
$\rho,\omega \to \pi^0\pi^0\gamma$, measurement of the
hadron production cross sections 
$e^+e^-\to 3\pi,\; 4\pi,\; \omega\pi,\; K\bar{K} $ \cite{talk1,talk2}. 
The study of the hadron cross sections will be continued
at VEPP-2000 \cite{vep2000}. This machine is being built in Novosibirsk
and will cover a poor studied energy region from
1.4 up to 2.0 GeV.

\begin{figure*}[t]
\centering
\includegraphics*[width=150mm]{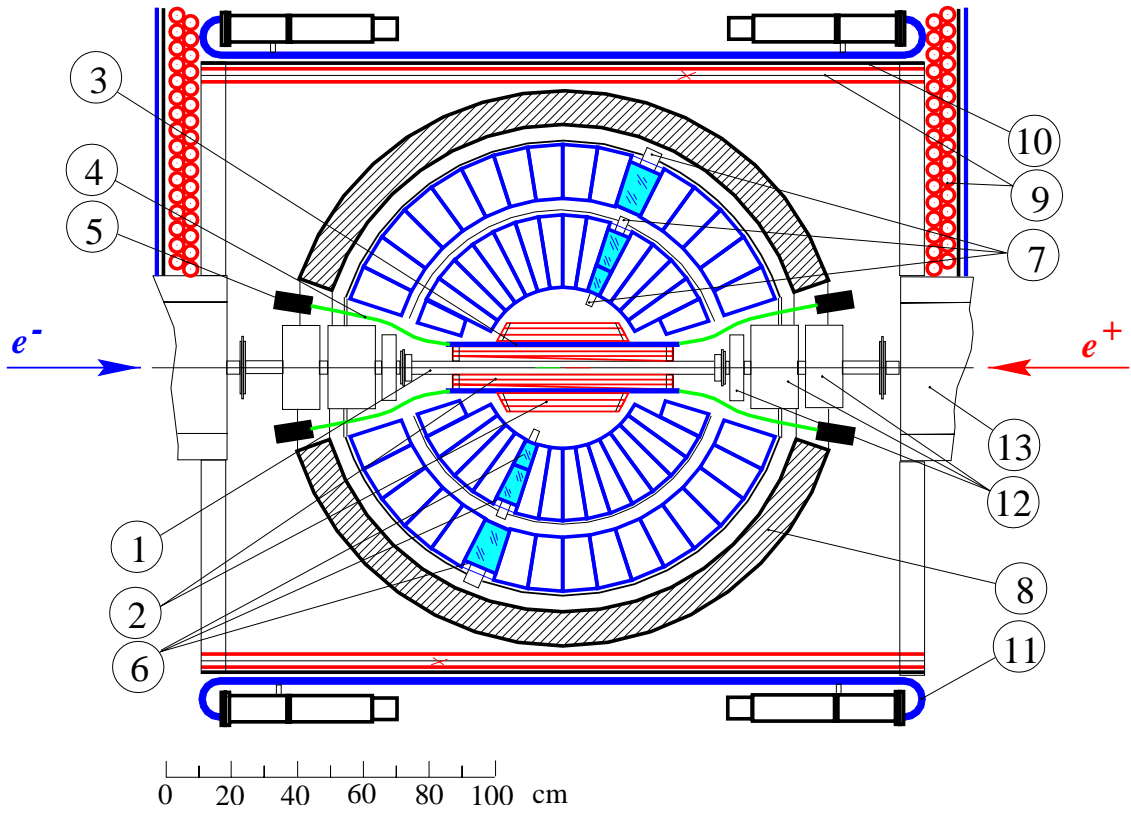}
\caption{SND detector layout: (1) beam pipe, (2) drift chambers,
(3) scintillation counters, (4) light guides, (5) PMTs,
(6) NaI(Tl) crystals, (7) vacuum phototriodes, (8) iron absorber,
(9) streamer tubes, (10) 1 cm iron plates, (11) scintillation counters,
(12) and (13) elements of collider magnetic system.}
\label{snd1}
\end{figure*}

The current layout of the detector is shown in Fig. \ref{snd1}.
The main part of SND is the electromagnetic calorimeter
based on NaI(Tl) crystals. Inside the calorimeter a tracking
system is placed. It consists of two drift chambers with a 
cylindrical scintillation counter between them. The calorimeter is
surrounded by a 12 cm thick iron absorber and a segmented muon system
which provides both muon identification and cosmic background
suppression. Each segment of this system consists of two layers 
of streamer tubes and a plastic scintillation counter. 

For experiments at VEPP-2000 a part of detector subsystems
will be modified.
\begin{itemize}
\item
The new design of the collider interaction region
requires a change of sizes of the tracking system.
The new
drift chamber is being designed now.
\item
The $dE/dx$ measurements in the
drift chamber can provide $\pi/K$ separation only up to 1200 MeV
center-of-mass energy. To cover completely the energy region of VEPP-2000
we plan to install an additional system for particle identification based
on aerogel \v{C}erenkov counters. 
\item 
The new phototriodes will be installed in the third calorimeter
layer. The old ones significantly degraded during experiments at
VEPP-2M.
\item
The high luminosity (up to $10^{32}$
cm$^{-2}$s$^{-1}$ at $E=$2 GeV) of the new machine and possible
increase of beam background require modification of the digitizing
electronics and new data acquisition system.
\item The combination of the relatively low energy and high luminosity
at VEPP-2000 gives a good possibility for study of $\gamma\gamma$
physics at low invariant masses $W_{\gamma\gamma}$. For this task,
detector will be equipped with a system for detection of the scattered
electrons.
\end{itemize}

\section{Calorimeter}

\begin{figure}[b]
\centering
\includegraphics*[width=70mm]{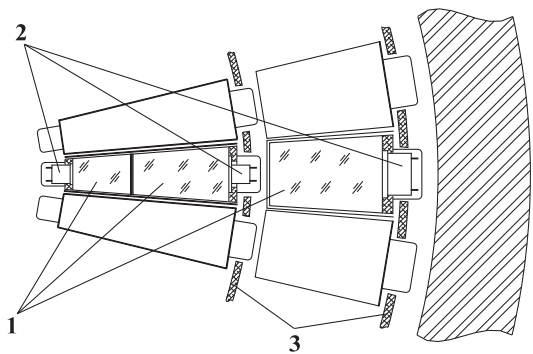}
\caption{NaI(Tl) crystals layout inside the calorimeter:
(1) NaI(Tl) crystals, (2) vacuum phototriodes, (3) aluminum supporting
hemispheres.}
\label{snd3}
\end{figure}

The calorimeter has a spherical shape which provides relative
uniformity of response over the whole solid angle. It consists of
1632 NaI(Tl)counters arranged in three layers. Pairs of counters of
the two inner layers are sealed in common 0.1 mm thick aluminum containers
and fixed to an aluminum supporting hemisphere (Fig.\ref{snd3}). Behind
it, the third layer is placed. The gap between the adjacent crystals of 
one layer is about 0.5 mm. 
The total calorimeter thick is 13.4 $X_0$
(34.7 cm) with 2.9, 4.8 and 5.7 $X_0$ in the first, second and third layers,
respectively. The total mass of NaI(Tl) is 3.5 t.

The calorimeter covers polar angle between 18 and 162 degrees.
The polar angle dimension of crystals is 9 degrees. The azimuthal angle dimension
is 9 degrees in ``large'' polar angle region $36^\circ<\theta<144^\circ$ and
18 degrees in the rest part. Each calorimeter layer contains crystals
of eight different shapes.
\begin{figure}[b]
\centering
\includegraphics*[width=75mm]{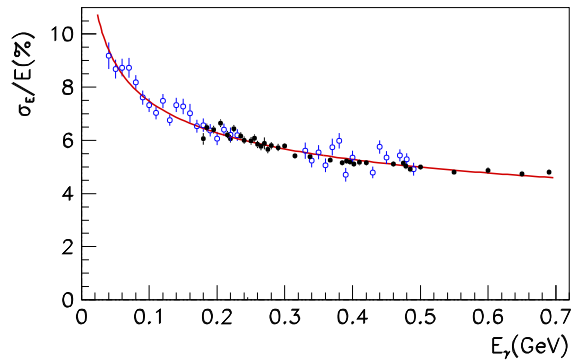}
\caption{Dependence of the calorimeter energy resolution on the photon
energy. The energy resolution was measured using $e^+e^-\to\gamma\gamma$
(dots) and $e^+e^-\to e^+e^- \gamma$ (circles) reactions.}
\label{snd7}
\end{figure}

The scintillation light signals from crystals are detected by vacuum
phototriodes. The light collection efficiency varies from
7\% to 15\% for crystals of different layers. The phototriode quantum
efficiency is about 15\% and the gain is about 10. 
The electronic
channel of the calorimeter consists of the charge sensitive
preamplifier, shaper and 12-bit ADC. The equivalent sensitivity of electronic
channel is about 0.25 MeV per ADC bit. The equivalent
noise lies within the 150-350 keV range.
\begin{figure}[t]
\centering
\includegraphics*[width=80mm]{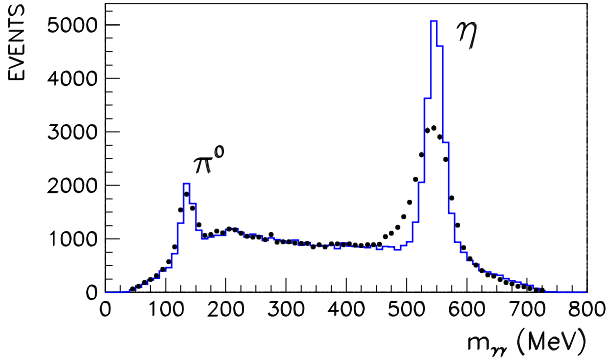}
\caption{Spectra of the invariant masses of photon pairs in
$e^+e^-\to\gamma\gamma\gamma$ events before (dots) and
after (line) kinematic fit.}
\label{snd11}
\end{figure}

The calorimeter energy resolution is determined mainly by the fluctuations
of the energy losses in passive materials before and inside the calorimeter,
leakage of shower energy through the calorimeter and nonuniformity of the
light collection. The most probable value of the energy deposition for photons
is about 93\% of their energy. The dependence of the calorimeter energy
resolution on photon energy is shown in Fig.\ref{snd7} and can be
approximated as:
$$ \sigma_E/E=4.2/\sqrt[4]{E(\mbox{GeV})}\%. $$

The calorimeter angular resolution on photon energy is described by
the following formula:
$$\sigma_\phi=0.82^\circ/\sqrt{E(\mbox{GeV})}\oplus 0.63^\circ.$$
The crystal width approximately matches the transverse size of 
an electromagnetic shower in NaI(Tl). Two showers can be distinguished
if the angle between them is larger then 9 degrees. If this angle
exceeds 18 degrees the energy of the showers can be measured separately 
without a loss of accuracy.

The two-photon invariant mass distribution of $e^+e^-\to 3\gamma$
process in the energy region of $\phi$ meson resonance is shown
in Fig.\ref{snd11}. The raw $\pi^0$ and $\eta$ mass resolutions are
11 MeV and 25 MeV, respectively. In the analysis of most part of
physical processes at SND the procedure of kinematic fitting is
used. This procedure allows to distinguish different processes
and improves the energy resolution. 
The effect of kinematic fit
to the invariant mass distribution is demonstrated in Fig.\ref{snd11}.
The improvement of mass resolution by a factor 1.5 is clearly seen.

The multilayer structure of calorimeter is used for $e/\pi$
\cite{pipi} and $K_L/\gamma$ \cite{ks3p} separation. 

\section{Tracking system}

\begin{figure}[t]
\centering
\includegraphics*[width=80mm]{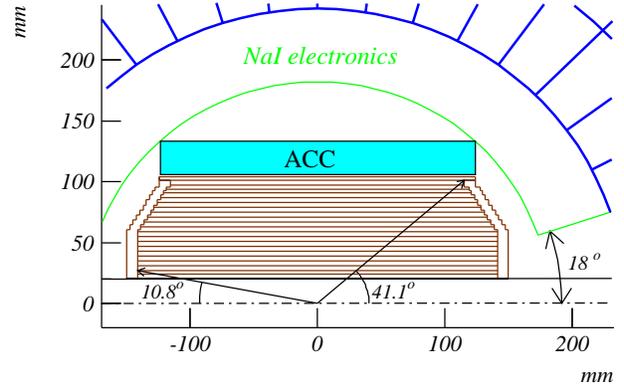}
\caption{Side view of the SND tracking system.}
\label{dcrz}
\end{figure}

\begin{figure}[t]
\centering
\includegraphics*[width=80mm]{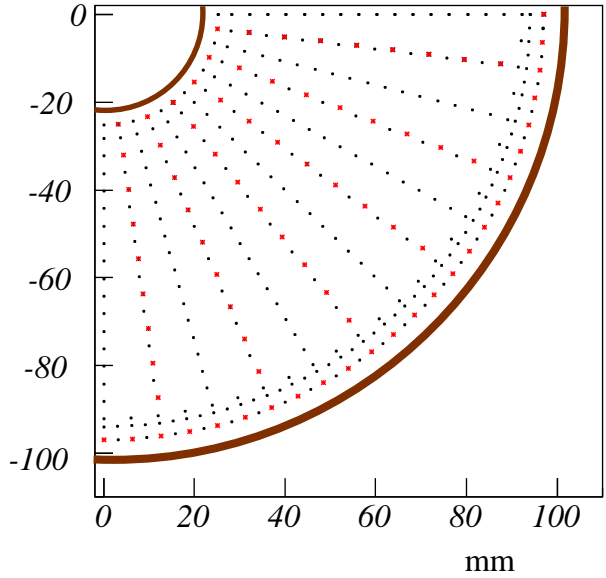}
\caption{SND tracking system view across the beam.}
\label{dcrphi}
\end{figure}

The new tracking system consists of a drift chamber and a proportional
chamber in the common gas volume (Figs.\ref{dcrz} and \ref{dcrphi}).
The gas volume has cylindrical shape with an inner radius of 2 cm and
an outer radius 10.3 cm. Its length varies from 30 cm to 26 cm.

Inside the gas volume there are 10 layers of sense wires.
The inner 9 layers is the drift chamber consisting
of 24 jet-type cells. The last outer layer is the proportional
chamber used for measurement of Z coordinate.
Both cylindrical walls have copper cathodes
divided into strips transverse to wires with the step of 6 mm
on their internal side.
Another side of the walls are used for cathode signal output.
The inner and outer cathodes have 128 and 152 strips, respectively.
Both chambers contain 312 sense and 984 field wires.
The sense wires are made of gold-plated tungsten
with a diameter of 15 $\mu$m. The field wires are 100 $\mu$m bronze-plated
titanium wires. In the drift chamber a $\pm$0.3 mm staggering of the sense 
wires in azimuthal direction is used to resolve the left/right ambiguity. 
The chamber will operate with a 90\%Ar + 10\%CO$_2$ gas mixture. 

The drift chamber electronics has to provide the drift time measurement
and the amplitude measurements from both ends of sense wire under
very hard background condition. The expected background rate of
individual wire is up to 50 kHz.
For time measurement, the front of the sum signal from both ends
of the wire is detected and the arrival time is digitized with 1 ns resolution.
The results of two time measurements during 1 $\mu$s trigger latency
is written to minimize the dead time. For amplitude measurement, 
a 10-bit 40 MHz FADC is used. The total charge can be obtained as
a sum of about 12 amplitude measurements.
The FADCs of the same kind are used to measure amplitudes of signals from
cathode strips.
\begin{figure*}[t]
\centering
\includegraphics*[width=150mm]{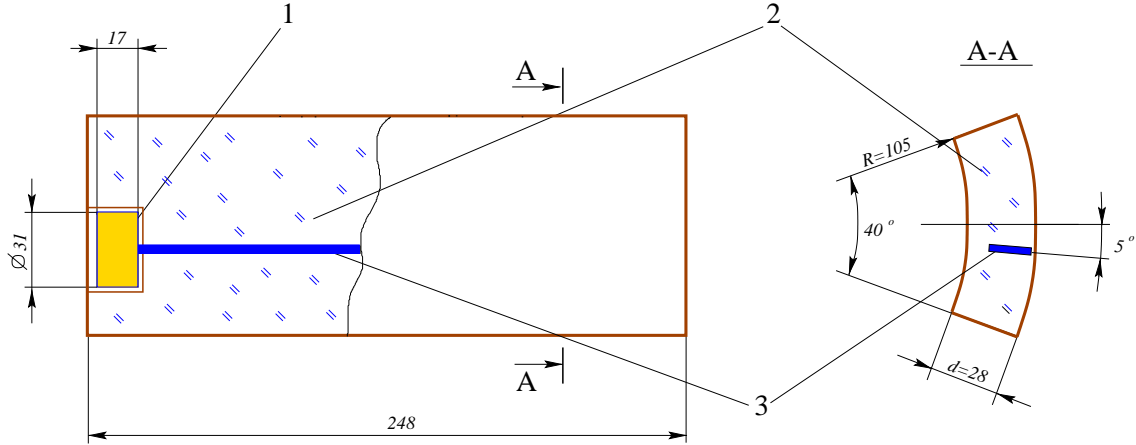}
\caption{The top view of SND aerogel \v{C}erenkov counter:
(1) photomultiplier, (2) aerogel radiator, (3) wave length shifter.}
\label{aero}
\end{figure*}

The expected drift chamber spatial resolution in azimuthal plane is
about 150 $\mu$m.  Z-coordinate will be measured by the charge division
method.  The $\sigma_z/l\sim0.5\%$ obtained in the test measurements with
40 MHz FADC corresponds to Z-resolution of 1-1.5 mm for different
chamber layers. The cathode strips provide 2 additional measurement
of Z with an expected accuracy of 0.3-0.6 mm depending on the track
polar angle. The overall track system angular resolution is 0.2
degrees in both azimuthal and polar directions.

The $dE/dx$ measurements in the drift chamber give a possibility
to separate $K$ and $\pi$ mesons with momenta up to 300 MeV/c.

\section{Particle identification system}

The SND identification system is based on \v{C}erenkov threshold counters and
should provide $K/\pi$ separation in the momentum region from 300 MeV/c to
870 MeV/c, which is maximal $K$ meson momentum at VEPP-2000. The aerogel with
relatively large refractive index equal to 1.13 was chosen as a
\v{C}erenkov light radiator. The corresponding threshold momenta for
$K$ and $\pi$ mesons are 246 and 938 MeV/c, respectively.

The identification system is a 25 cm long cylinder surrounding the tracking 
system (Fig.\ref{dcrz}). The inner cylinder radius is 10.5 cm. The width
is 2.8 cm. The system consists of 9 equal counters with an azimuthal size
of 40 degree (Fig.\ref{aero}). For the \v{C}erenkov light collection, the wave length shifters
(WLS) are used \cite{onuchin}. This method allows to diminish a number of 
the detector channels but leads to a possibility of misidentification for 
particles
struck into WLS. In SND case the WLS azimuthal size is 3 mm and
possibility of the particle misidentification is about 5\%.
The WLS light is detected by a compact photomultipliers with 
microchannel plates. For SND geometry, the expected number of the 
photoelectrons from an ultrarelativistic particle is about 10.

\section{Electron tagging system}

\begin{figure*}[t]
\centering
\includegraphics*[width=130mm]{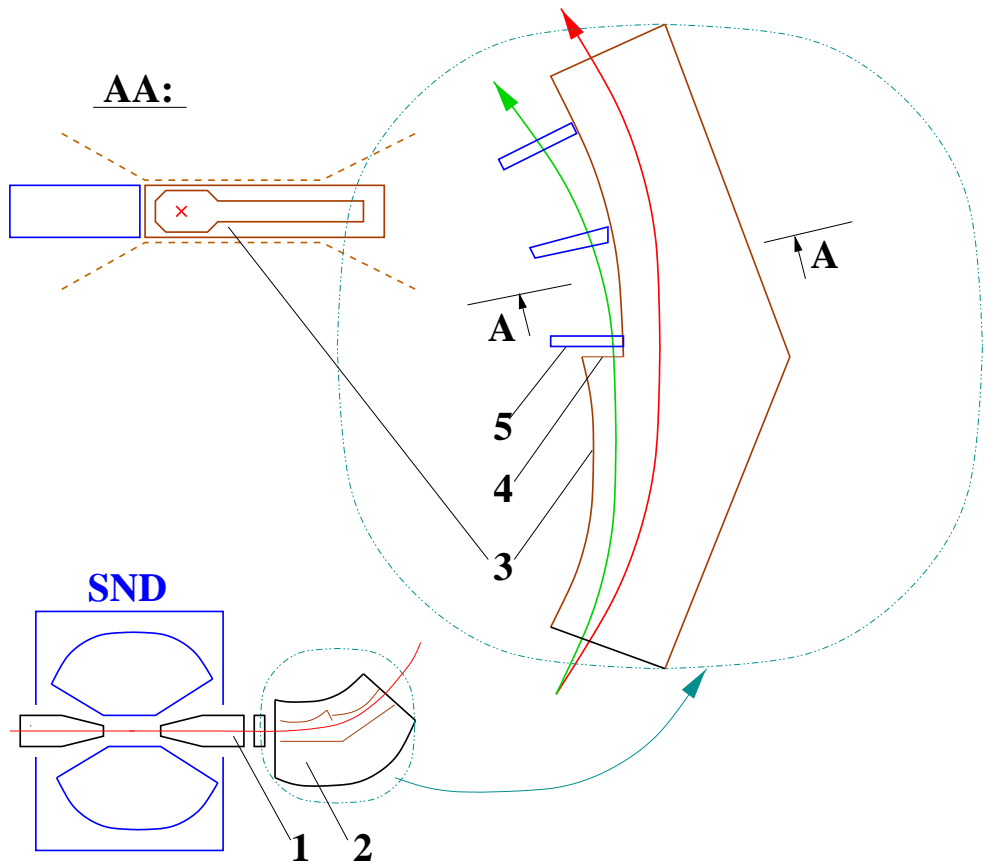}
\caption{The view of SND electron tagging system:
(1) superconducting solenoid, (2) bending magnet, (3) vacuum pipe,
(4) window, (5) electron detector. }
\label{tagsys}
\end{figure*}

The VEPP-2000 gives a good possibility for study of $\gamma\gamma$
physics at low $\gamma\gamma$ invariant mass.  The SND detection
efficiency is about 50\% for $\gamma\gamma\to \pi^0,\, \eta,\, \pi^0\pi^0$
processes. The integrated luminosity of 150 pb$^{-1}$ at 2 GeV (1 month at
expected luminosity of $10^{32}$ cm$^{-2}$sec$^{-1}$) corresponds to
$7.5\cdot10^4 \pi^0$, $2.5\cdot10^4 \eta$, $1.8\cdot10^4 \eta^\prime$,
2000 $f_0$, 1000 $a_0$ produced in the photon-photon collisions.
To successfully select two photon events from the beam background
and the events of $e^+e^-$ annihilation we plan to install the
system for detection of the scattered electrons.  The project
of the electron tagging system is under study now. The one of possible
designs is shown in Fig.\ref{tagsys}. The two
set of the electron detectors are placed inside the bending magnets
of the collider near the interaction region. The electron detector can
be a GEM (gas electron multiplier) based detector with two coordinate
readout. The presented system can provide the double tag efficiency
of 15-25\% in the $\gamma\gamma$ invariant mass region from 300 to 1000 GeV
and the mass resolution of about 5 MeV.

\section{Data acquisition system}

The SND data acquisition system is based on KLUKVA electronics
standard \cite{klukva} developed at BINP. One KLUKVA crate
holds up to 16 data conversion (DC) modules, first level trigger interface
(IFLT) and readout processor (RP) modules. The DC modules
provide fast logical and analog signals, which are transfered via KLUKVA 
bus to IFLT modules for further use in FLT. The digitized data are
extracted from the DC modules by RP with 100 ns cycle.
The SND digitizing electronics occupies 16 KLUKVA crates.

The first level trigger system uses the following information:
\begin{itemize}
\item total energy deposition in calorimeter,
\item 160 logical signals from calorimeter towers (25 MeV threshold);
\item 216 logical signals from drift chamber wires.
\end{itemize}
The FLT logic is implemented as a pipe line working at 40 MHz clock rate.
The FLT decision latency is about 1 $\mu$s. The FLT rate in
the experiments at VEPP-2000 is expected to be about 1000 Hz.

The most part of KLUKVA modules including RP will be modified for
new experiments. The pedestal subtraction and zero suppression will
be performed in the DC modules. As a result an average event size
will be about 3 kB. The time needed for this event digitization and
reading into RP modules is about 30 $\mu$s. New RP module provides
data exchange with online computer via Ethernet network.
\begin{figure}[t]
\centering
\includegraphics*[width=80mm]{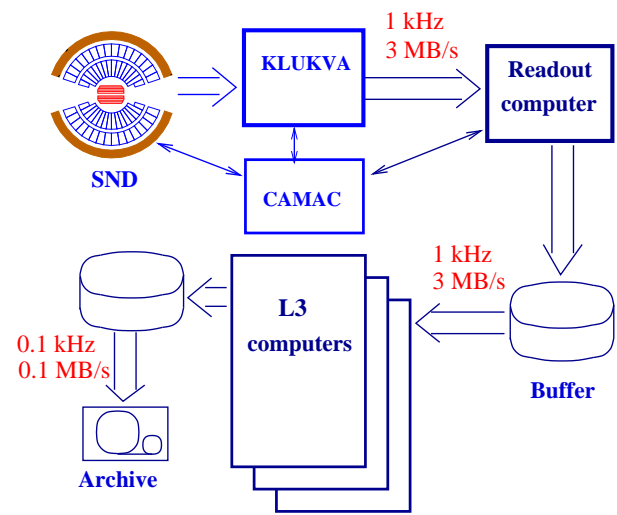}
\caption{SND data acquisition system.}
\label{online}
\end{figure}

A schematic view of the SND data acquisition system is shown in
Fig.\ref{online}. The raw SND data are read by the Readout computer.
The average data flow at this stage  is about 3 MB/s.
The Readout computer builds events and writes data into buffers
on SCSI disks shared by several computers.
The L3 processes (4 or more) taking data from
buffers provide event packing, partial reconstruction, filtering and
recording.  This DAQ structure allows to increase an event processing
rate proportionally to the number of L3 processes. The resulting output 
data flow is expected to be about 0.1 MB/s.

\section{Offline software}
Present FORTRAN-based offline programs will be replaced with
object-oriented framework, which supports simulation, reconstruction
and analysis activities.  New framework exploits the experience
obtained in the work with the current offline and 
supports or extends their essential features.
Similar projects that exists in HEP (e.g. BaBar framework)
were also examined to find out possible weak and strong features.

The main framework concept is a module, which is basic processing unit
consuming some data and producing more data.  Every module can be parameterized
during run time.  Formalized description of modules is used by framework
sequencer for the selection and ordering of minimal subset of modules for
any given task.  Data persistency services are made sufficiently abstract to
allow implementation for different
persistency technologies.  The framework provides an interface for
scripting languages.  Together with a custom expression parser this gives a
support for extensible run-time histogramming.

\section{Conclusion}
For experiment at VEPP-2000 a part of SND subsystems will be
upgraded: the electromagnetic calorimeter, tracking system,
electronics, online and offline software. Two new subsystems will
be added to detector: aerogel \v{C}erenkov counter and electron
tagging system. All these changes significantly extend SND capabilities
for studies of the physical processes in new energy domain.

\end{document}